\documentstyle[aps,prl,twocolumn,epsfig,epsf,floats]{revtex}

\begin{document}
\wideabs{
\author{F. M. Cucchietti$^{1}$, H. M. Pastawski$^{\ast,1,2}$ and D. A. Wisniacki$^{3}$}

\address{$^{1}$ {\it Facultad de Matem\'{a}tica, Astronom\'{\i}a y F\'{\i}sica,Universidad Nacional de C\'{o}rdoba,}\\
{\it Ciudad Universitaria, 5000, C\'{o}rdoba,Argentina}}
\address{$^{2}$ {\it Institut de Physique et Chimie des Mat\'{e}riaux de Strasbourg, UMR 7504,}\\
{\it CNRS-ULP, 23 rue du Loess, 67037 Strasbourg Cedex, France} }
\address{$^{3}$ {\it Departamento de F\'{\i}sica, Comisi\'{o}n Nacional de Energ\'{\i}a At\'{o}mica,}\\
{\it Avenida del Libertador 8250, 1429 Buenos Aires, Argentina} }

\title{Decoherence as Decay of the Loschmidt Echo in a Lorentz Gas.}


\date{January 29 2001}

\maketitle

\begin{abstract}
Classical chaotic dynamics is characterized by the exponential sensitivity
to initial conditions. Quantum mechanics, however, does not show this
feature. We consider instead the sensitivity of quantum evolution to
perturbations in the Hamiltonian. This is observed as an atenuation of the
Loschmidt Echo, $M(t)$, i.e. the amount of the original state 
(wave packet of width $\sigma$) which is
recovered after a time reversed evolution, in presence of a classically weak
perturbation. By considering a Lorentz gas of size $L$, which for large $L$ 
is a model for an {\it unbounded} classically chaotic system, we find numerical evidence that, if the
perturbation is within a certain range, $M(t)$ decays exponentially with a
rate $1/\tau _{\phi }$ determined by the Lyapunov exponent $\lambda $ \ of
the corresponding classical dynamics. This exponential decay extends much
beyond the Eherenfest time $t_{E}$ and saturates at a time $t_{s}\simeq \lambda^{-1}\ln (\widetilde{N})$, where $\widetilde{N}\simeq (L/\sigma)^2$ is the effective dimensionality of the Hilbert space. Since $\tau _{\phi }$ quantifies the
increasing uncontrollability of the quantum phase (decoherence) its characterization and control has fundamental interest.

\end{abstract}

}
Chaos justifies the observed \ macroscopic irreversibility within the
reversible laws of Classical Mechanics. One of its characteristic features
is the exponential divergence of trajectories corresponding to nearby
initial conditions, which leads to deterministic unpredictability. However,
quantum dynamics exhibits insensitivity to initial conditions \cite
{cit-QChaos-europ} and hence prevents a dynamical definition of quantum
chaos. Therefore, quantum signatures of chaos in systems with a chaotic
classical equivalent are searched in their steady-state properties \cite
{cit-Gutz} such as spectral rigidity\cite{cit-Bohigas}, wave function
morphologies\cite{cit-scars}, and decaying parametric correlations of
observables \cite{cit-Aaron}. Early attempts \cite{cit-Peres} to address
unitary quantum dynamics did not clarify the connections with the dynamical
classical concepts of chaos, but the inclusion of interactions with a
dissipative environment was expected \cite{cit-Zurek-Paz} to produce an
entropy growth controlled by chaos. In a purely Hamiltonian problem, quantum
reversibility can be monitored\ through the amount of revival of a local
density excitation, upon time reversal of its unitary quantum evolution,
i.e. the Loschmidt Echo\cite{cit-loschmidt}. By considering a Lorentz gas
where the reversed evolution is disturbed by a static perturbation, we find a
Loschmidt Echo that attenuates exponentially with a rate associated to the
chaos of the classical system. Under appropriate conditions, the\ dynamical
instability of the classical system translates into quantum phase
unpredictability (decoherence) and the classical Lyapunov exponent becomes
the quantum irreversibility rate.

A hypersensitivity of time reversal to perturbations was observed in recent
NMR experiments on many-body spin systems\cite{cit-REPE}\cite{cit-Potencia}.
In essence, these experiments\cite{cit-polecho}\cite{cit-mesosc-echos}
involve the creation of a local density excitation represented by a state $%
\left| 0\right\rangle $ which evolves under a many-spin Hamiltonian 
${\cal H}.$ The Loschmidt Echo (LE) is the probability to return to the initial state
when a Hamiltonian evolution for a time $t$ is followed by an identical
period of imperfect reversal of that evolution, achieved by the
transformation ${\cal H\rightarrow -}({\cal H}+\Sigma )$, i.e.: 
\begin{equation}
M(t)=|\left\langle 0\right| \exp [{\rm i}({\cal H}+\Sigma )t/\hbar ]\exp [-%
{\rm i}{\cal H}t/\hbar ]\left| 0\right\rangle |^{2},  \label{eq-Overlap}
\end{equation}
where $\Sigma $ is a constant (or quasi-static) Hermitian perturbation
representing the imperfection in the Hamiltonian reversal. Notice that in
these experiments $-\ln [M(t)]$ is a measure \cite{cit-REPE} of the growing
entropy. A striking finding \cite{cit-Potencia} was that, for small $\Sigma $%
, the {\it decay} of $M(t)$ becomes independent of $\Sigma $, being
proportional only to the dynamical scales of ${\cal H}$. The hints that
chaotic systems may be unstable under fluctuating perturbations\cite
{cit-Zurek-Paz} and that many-body systems could be assumed to be
intrinsically chaotic \cite{cit-Gaspard-NATURE} suggested us the hypothesis
that the most relevant parameters of the Hamiltonian dynamics, the Lyapunov
exponents $\lambda $, might control the decay of $M(t)$. Moreover, a
semiclassical analysis, assuming \cite{cit-Jalabert-Past} a one-body
classically chaotic Hamiltonian with a perturbative potential representing a
quenched disorder predicted a LE decaying exponentially with a rate $1/\tau
_{\phi }=\lambda .$ These predictions, based on various approximations,
encouraged us to perform exhaustive numerical experiments.

\begin{figure}[tb]
\centering \leavevmode
\center{\epsfig{file=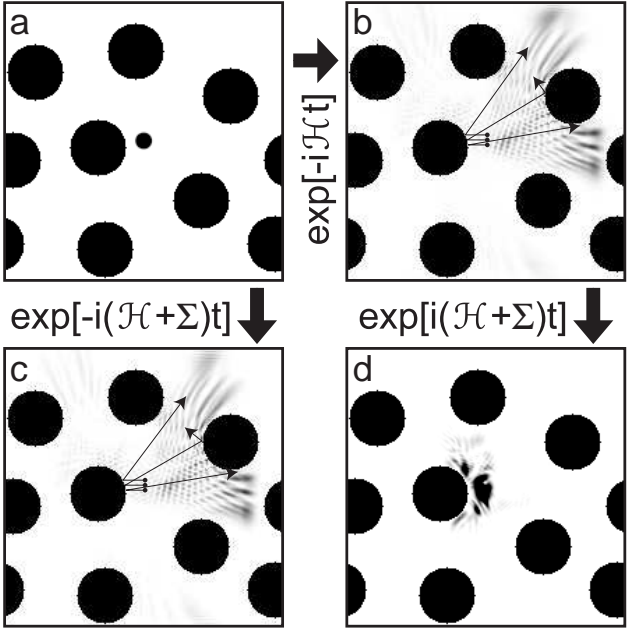, ,width=8.3cm,angle=0}}
\vspace{0.5cm}
\caption{Wave packet evolution in a system with $L/a=200$, $R/a=20$ and $c=8%
\protect\pi \times 10^{-2}$. a) Initial state at $t=0$. The velocity points
to the left. b) State evolved with the unperturbed Hamiltonian for a time $%
t=30\hbar /V$. c) State evolved with the perturbed Hamiltonian ($\protect%
\alpha =0.04$) for a time $t=30\hbar /V$. d) State evolved from that
depicted in panel b) with the perturbed Hamiltonian for a time $t=-30\hbar
/V $. The square of the overlap (Loschmidt Echo) between the states of
panels a) and d) is $M(t)=0.09$, same as that between the states of panels
b) and c).}
\label{fig-paneles}
\end{figure}

We consider a Lorentz gas, i.e.: a particle in a square billiard of area $%
L^{2}$ where we fix an irregular array of $N$ circular scatterers
(impurities) of radius $R$. A particular realization of such system is
represented in Fig. \ref{fig-paneles}-a where the length, in a minimal unit $%
a$, is $L=200a.$ The scatterers concentration is $c=N\pi R^{2}/L^{2}=8\pi \times
10^{-2} $. The minimum distance allowed between two scatterer centers is $3R$%
. This requirement, together with the imposed periodic boundary conditions,
prevents the geometrical localization. 

The Lyapunov exponent of a Lorentz
gas should be: $\lambda =\beta v/\ell $, where $\ell $ is the collision mean
free path, $v$ the particle velocity and $\beta \sim 1$ a geometrical factor 
\cite{cit-Laughlin} depending logarithmically on $\ell $. In our system, we estimate $\ell \simeq L^{2}/(2NR)-\pi
R/2.$ The computation of the distribution of distances between collisions
gives a shifted Poisson distribution whose mean value is $\ell $. The
Lyapunov exponent is obtained \cite{cit-Lyapunov} \ from the average
logarithm of the distance $d_{t},$ after a time $t,$ between two classical
trajectories initially separated by a small distance $d_{0}$. The condition
of smallness for $d_{0}$ is that $d_{t}\ll R.$ The longer the time the more
precise the estimation of $\lambda $. By neglecting correlations in the
position of the impurities, we obtain the estimation:

\begin{equation}
\beta \simeq \int_{R}^{\infty }{\rm d}s\frac{\exp (-s/\ell )}{\ell \exp
(-R/\ell )}\int_{0}^{1}{\rm d}x\ln [1+\frac{s}{R\sqrt{1-x^{2}}}].
\label{eq-lyap-theor}
\end{equation}
The outer integral accounts for the distribution of free paths, $s$, and the
internal one is the distribution of impact parameters $x$. Numerical $%
\lambda \ell /v$ verify Eq. (\ref{eq-lyap-theor}). For the present case of $%
c=8\pi \times 10^{-2}$ , i.e. $\ell =96a,$ we got $\beta $ =$(2.1\pm 0.1).$

The perturbation parameter $\alpha $ controls the distortion of the diagonal
components of the mass tensor , $m_{x,x}=m_{o}(1+\alpha )$, and $%
m_{y,y}=m_{o}/(1+\alpha )$, with $m_{o}$ the isotropic unperturbed mass.
This perturbation is inspired in the effects of a slight rotation of the
sample in the related problem of dipolar spin dynamics \cite{cit-QZeno},
which modifies the mass of the spin wave excitations. It is written as

\begin{equation}
\Sigma (\alpha )=-\frac{\alpha }{1+\alpha }\frac{p_{x}^{2}}{2m_{o}}+\alpha 
\frac{p_{y}^{2}}{2m_{o}}.  \label{eq-SIGMA-mass}
\end{equation}

For a fixed initial position and velocity we find numerically that two
evolutions with slightly different mass tensors lead to trajectories whose
distance in phase space grows exponentially with the same Lyapunov exponent
that amplifies initial distances: i.e. the classical dynamics of this system
is equally sensitive to changes in the Hamiltonian as to changes in the
initial conditions\cite{cit-Shack-Caves}.

The Hamiltonian operator is obtained by a lattice discretization on a small
scale $a$. The evolution is calculated using the Trotter-Suzuki (TS)
algorithm\cite{cit-DeRaedt}. Its basic idea is that, by choosing a short
time unit $\tau $, the evolution can be approximated as the finite product
of evolution operators where each is solved analytically. In the lowest
order, 
\begin{equation}
U(\tau )=\exp [{\rm i}{\cal H}\tau /\hbar ]\simeq \widetilde{U}(\tau
)=\prod_{k}^{Q}\exp [{\rm i}{\cal H}_{k}\tau /\hbar ],
\label{eq-TROTTER-evol}
\end{equation}
where ${\cal H}=\sum_{k}^{Q}{\cal H}_{k}.$ The conceptual virtue of this
method is that $\widetilde{U}(\tau )$ is always unitary. For an
approximation of order $n$, the difference between $\widetilde{U}(\tau )$
and $U(\tau )$ is order $\tau ^{n+1}$. Our choice of $n=4$ and $\tau
=0.1\hbar /V=0.2m_{o}a^{2}/\hbar $ allowed us to obtain high accuracy even
for times an order of magnitude larger than needed.

\begin{figure}[tb]
\centering \leavevmode
\center{\epsfig{file=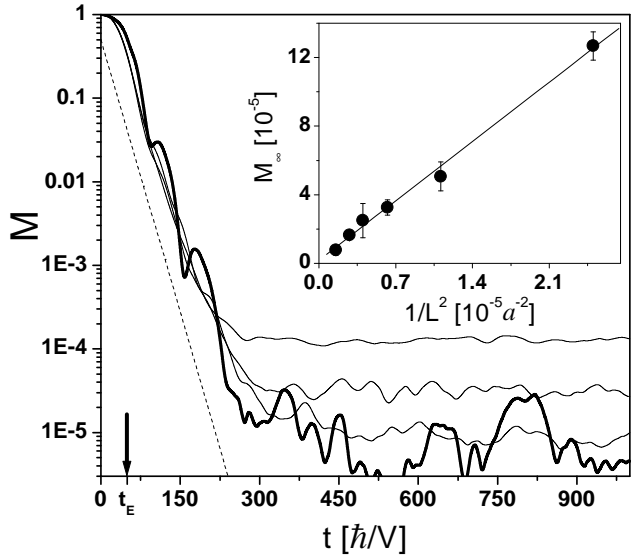,width=8.3cm}}
\vspace{0.5cm}
\caption{In bold line, $M(t)$ for an individual system with $L/a=800$ and $%
\protect\alpha =0.024$. The Ehrenfest time is shown with an arrow. In dashed line, shown for reference, exponential
decay with the calculated Lyapunov exponent of the classical equivalent. In
solid lines, average $M(t)$ for different system sizes $L/a=$200, 400 and
800 and the same perturbation. In the inset $M_{\infty }$ is shown for $%
L/a=200$, 300, 400, 500, 600 and 800 as a function of $(a/L)^{2}$. The
straight line is the best fitting, with $M_{\infty }=(5\pm 0.2)(a/L)^{2}$}
\label{fig-minfinito}
\end{figure}

Let's consider a typical system with $R/a=20$, $c=8\pi \times 10^{-2}$
and size $L/a=200$ $.$ The initial state is a Gaussian wave-packet of width $%
\sigma =3a$ and wave number $k_{x}=-0.7/a$, $k_{y}=0$ $\ $shown in Fig. \ref
{fig-paneles}-a.\ All further simulations will use $k_{x}=k_{y}=3\pi /8a$.
Since $kR\gg 1$, comparison with semiclassical calculations is justified.
We took $2\pi /k\ll \ell =(96\pm 2)a$, in order to minimize Anderson's 
localization effects. The density resulting from a typical evolution for a time $%
t=30\hbar /V$ \cite{cit-website} is shown in panel b)$.$ An evolution with
the perturbed Hamiltonian ($\alpha =0.04$) for the same time is shown in
Fig. \ref{fig-paneles}-c. In both panels the classical trajectories
corresponding to three initial positions are shown for reference. While
densities look similar to the eye, $M(t)$ is about 0.09 indicating the
relevant role of the quantum phase in the attenuation of $M(t)$. Panel d)
shows the LE formed by the perturbed backward evolution of the state in
panel b). Analyzing $M(t)$ in different realizations we observe that, after
the initial transient, $M(t)$ fluctuates around an exponential decay, with a
characteristic time $\tau _{\phi }.$ Notably this exponential decay persists
much beyond the Ehrenfest time $t_{E}\simeq \frac{1}{\lambda }\ln
[2R/\lambda _{F}]\simeq 40$ $\hbar /V$, which is fixed by the local scale of
the potential fluctuation\cite{cit-Zurek-Paz}. In a Lorentz gas, in contrast
with the usual case of chaotic cavities, $t_{E}$ is independent of the
system size\cite{cit-Aleiner-Larkin}. Finally, $M(t)$  saturates in a time
scale related to system size. A typical curve for $L/a=800$ is shown with a
bold line in Fig.(\ref{fig-minfinito}). 

In principle, any given precision of 
$\tau _{\phi }$ can then be obtained studying a system large enough, however
this easily becomes computationally expensive.
Alternatively, one can resort to the ensemble averaging of the observable $%
M(t),$ which reduces noise and defines $\tau _{\phi }$ with the same
precision in much smaller systems. This is exemplified in Fig. \ref
{fig-minfinito}, where we present the ensemble average $M(t)$ for billiards
of three different sizes of $L/a=$200, 400 and 800 and fixed $\alpha =0.024$
and $c=8\pi \times 10^{-2}$ which show the same exponential behavior with a
progressively expanded range. Similar plots are obtained for the other
parameters with an exponential $M(t)$, showing that the ensemble average
does not introduce any spurious effect in the decay. Previous attempts to
characterize quantum chaos \cite{cit-Peres} considered the limiting value $%
M_{\infty }$ for the Loschmidt Echo at asymptotically large times. It was
proposed \ $M_{\infty }\propto 1/\tilde{N}$, where $\tilde{N}$ is  the
number of energy levels appreciably represented in the initial state. In our
case $\tilde{N}\simeq \left( L/\sigma \right) ^{2}$. The numerical results
verify a complete independence of $M_{\infty }$ on the perturbation. The
calculated $M_{\infty }$ for various sizes are shown in the inset of Fig. (%
\ref{fig-minfinito}) for a fixed $\alpha =0.024$ . The predicted
relationship is shown with a straight line obtaining $M_{\infty }=(5.0\pm
0.1)(a/L)^{2}$. For all the sizes we kept $c$ (and hence $\ell $, $\lambda$ and $t_E$) fixed and
we verified that $\tau _{\phi }$ does not depend on $L/a$. These results
imply that the Lyapunov exponential behavior persists up to a time $%
t_{s}\simeq \lambda ^{-1}\ln [\tilde{N}].$

\begin{figure}[tb]
\centering \leavevmode
\center{\epsfig{file=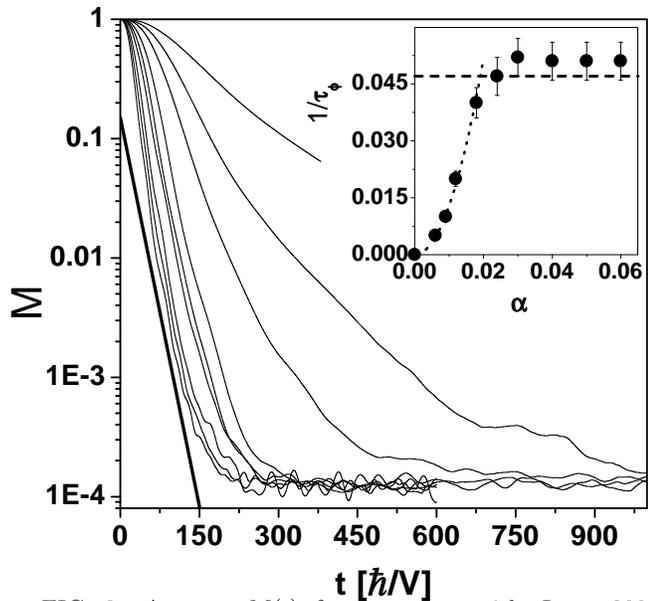, width=8.3cm}}
\vspace{0.5cm}
\caption{Average $M(t)$ for a system with $L=200a$, $R=20a$, which makes $%
\ell =(96\pm 2)a$. The values of $\protect\alpha $ are, from top to bottom:
0.006, 0.009, 0.012, 0.018, 0.024, 0.03, 0.04, 0.05 and 0.06. The thick line
corresponds to an exponential with the calculated numerical Lyapunov
exponent of the system. For long times $M(t)$ saturates to a finite value, $%
M_{\infty }$. In the inset the inverse characteristic decay times are shown,
evidencing the regime where the decay of M is given by the intrinsic
properties of the system. The dotted is the fit, $\protect\tau _{\protect\phi
}^{-1}=(130 \pm 6) \protect\alpha^{2} V/\hbar$, while the dashed line is the
classical Lyapunov exponent.}
\label{fig-Overlapdirecto}
\end{figure}

A representative dependence of $\tau _{\phi }$ \ on the parameter $\alpha $ is
obtained considering the smallest sample size compatible with a good
observation of the exponential, in this case $L/a=200.$ Fig. \ref
{fig-Overlapdirecto} shows $M(t)$ averaged over $100$ realizations that
contain at least a scatterer in the classical trajectory of the wave packet.
For $\alpha >\alpha _{c}\simeq 0.02$, all the exponential decays coincide, within
the numerical error, with the Lyapunov decay associated with the classical
system, shown with a thick line for comparison. The initial perturbative
parabolic decay, $M(t)\simeq 1-b$ $\left( \alpha t\right) ^{2}$ with $%
b\simeq 0.37\left( V/\hbar \right) ^{2}$, has a characteristic scale which 
{\it does} depend quadratically on the perturbation strength and prevents
the curves to lie superimposed. We show in the inset of Fig. \ref
{fig-Overlapdirecto} the numerical values of $\tau _{\phi }^{-1}$ {\it %
extracted from the exponential part}. It shows an initial quadratic
dependence on $\alpha $ and a cross-over at $\alpha _{c}\simeq 0.02$ to a
saturation at a value close to the classical Lyapunov exponent.

The semiclassical analysis predicts a universal behavior provided that the
perturbation is strong enough to be quantically significant, but weak enough
to be classically irrelevant. The first condition implies that the length $%
\widetilde{\ell }=v\widetilde{\tau }$ required for an important change in
the phase as consequence of the perturbation must be shorter than the
distance associated to the Lyapunov exponent\ $v/\lambda $. A rough
estimation for our perturbation $\delta m_{x,x}=\alpha m_{o}$ is: 
\begin{equation}
\widetilde{\tau }\simeq \frac{\pi \hbar }{\delta E_{k}}=\frac{\pi m_{o}a^{2}%
}{\alpha \hbar (1-\cos [ka])}\lesssim 1/\lambda  \label{eq-Energy-change}
\end{equation}
which would predict a critical value $\alpha _{c}\simeq 0.13.$ However, we
see the universal exponential even for perturbations $\alpha \geq 0.02$
implying a critical value an order of magnitude smaller than Eq. (\ref
{eq-Energy-change}). This suggests that either this estimation or the
condition required in Ref. \cite{cit-Jalabert-Past} is too strong\cite
{cit-Jacquod-FGR}.

The condition for a classically weak perturbation means that it must not
modify the system's global properties. Otherwise, even in absence of chaos
(no collisions) the perturbation would spread out the classical
trajectories. In our case, this $M(t)$ fits a Gaussian decay. If at the time 
$1/\lambda $, required for the chaos to set in, the overlap has already
decayed by cause of the perturbation, the exponential decay would not be
observed. This sets an upper bound for the perturbation \ from $%
M(t=1/\lambda )=\exp [-b(\alpha /\lambda )^{2}]\leq M_{\infty }$. Replacing
with the parameters of the system shown in Fig. (\ref{fig-Overlapdirecto})
one gets the condition $\alpha \leq 0.3$, consistent with our results.

Therefore, we have shown that in a wide range of parameters $M(t)$ in any
large enough individual system decays exponentially until an asymptotic
value is reached. The characteristic decay time does not depend on the
perturbation, but rather on the intrinsic properties of the Hamiltonian. The
range of perturbation parameters with an exponential decay results to be
much broader than hinted by the previous semiclassical analysis. From our
numerical results one can infer that\ infinite systems not showing
Anderson's localization should present an unbounded exponential decay. The
Lorentz gas represents a broad class of chaotic systems, where it is
expected the same overall behavior. It has the additional advantage over 
other chaotic cavities that here the saturation time ($t_{s}\simeq \frac{2}{\lambda }\ln
[L/\sigma ]$) can be made arbitrarily longer than the Ehrenfest time. This
manifests the quantum nature of the observed effect. The simplicity of our
perturbation evidences a very remarkable property: {\it there is no need of
chaoticity or stochasticity in the ``environment'' to introduce
irreversibility}. Fig. \ref{fig-paneles} manifests that the irreversibility
produced by small perturbations results through their effect on the wave
function's phase. In this sense, the definition of classical chaos in terms
of hypersensitivity to initial conditions and perturbations translates, in
the quantum world, as a sensitivity of phases to perturbations in the
Hamiltonian, i.e. decoherence. Since we restricted ourselves to a
Hamiltonian problem, its solution clarifies how chaos limits our control of
dynamics even within the reversible framework of Quantum Mechanics of closed
systems. This concept has deep consequences for the diverse fields where
decoherence must be minimized, such as classical waves propagation \cite
{cit-waves}, mesoscopic and molecular electronics \cite{cit-Mol-Elect-review}
and quantum information\cite{cit-Quant-Inf-review}. In view of the entropic
meaning of $-\ln [M(t)]$ the interest for foundations of Statistical Physics 
\cite{cit-Lebowitz} is clear.

In summary, we have presented numerical evidence supporting the fact that
decoherence, as measured by the exponential decay of the Loschmidt Echo, is
controlled by the same parameters that govern classical chaos. This renders $%
M(t)$, defined in Eq. \ref{eq-Overlap}, as a very practical tool to study
dynamical quantum chaos.

We acknowledge the Argentinian Supercomputer Centre and the UNC Department
of Computer Science. DAW and FMC receive support from CONICET and SeCyT-UNC
respectively. HMP and PRL are affiliated to CONICET. This work received
financial support from CONICET, ANPCyT, SeCyT-UNC, Fundaci\'{o}n Antorchas
and the french-argentinian ECOS-Sud program.

$^{\ast }$Correspondent author. E-mail: horacio@famaf.unc.edu.ar

\end{document}